\documentclass[oneside]{amsart}

\usepackage[english]{babel}
\usepackage{bbm,euscript,amssymb,amsthm}
\usepackage{mathrsfs}
\usepackage{amsfonts}
\usepackage{amsmath}
\usepackage{amssymb}
\usepackage{fancybox}
\usepackage{graphicx}
\usepackage{color}

\usepackage{lineno}

\usepackage{pstricks,pstricks-add,pst-math,pst-xkey}

\usepackage{rotating}

\definecolor{darkgreen}{rgb}{0,.5,0}

\newtheorem{thm}{Theorem} 
\newtheorem{thm*}{Theorem*}
 
\newtheorem{theorem}[thm]{Theorem}

\newtheorem{observation}[thm]{Observation}
\newtheorem{corollary}[thm]{Corollary}

\def\eucl#1#2{{\Vert#1 - #2\Vert}}

\def\ignore#1{{ }}

\def\Pp{{\{P^t\ | \ t\in [a,b]\}}}
\def\Pp#1#2{{P^{[#1,#2]}}}

\def\ignore#1{}

\def\pp{{\mathscr P}}

\title{Visibility-preserving convexifications using single-vertex
  moves}

\author{B.M.~\'Abrego}
\address{Department of Mathematics, California State University. Northridge CA.}
\email{{\tt bernardo.abrego@csun.edu}}

\author{M.~Cetina}
\address{Instituto de F\'\i sica, UASLP. San Luis Potosi, Mexico.}
\email{mcetina@ifisica.uaslp.mx}

\author{J.~Lea\~nos}
\address{Unidad Acad\'emica de Matem\'aticas, UAZ. Zacatecas, Mexico.}
\email{jleanos@mate.reduaz.mx}

\author{G.~Salazar}
\address{Instituto de F\'\i sica, UASLP. San Luis Potosi, Mexico.}
\email{gsalazar@ifisica.uaslp.mx}
\thanks{Supported by CONACYT grant 106432.}

\date{\today}



\begin{document}


\begin{abstract}
Devadoss asked: (1) can every polygon be convexified so that no 
internal visibility (between vertices) is lost in the
process? Moreover, (2) does such a convexification exist, in which
exactly one vertex is moved at a time (that is, using {\em
  single-vertex moves})? We prove the redundancy of the ``single-vertex
moves'' condition: an affirmative answer to (1) implies an affirmative
answer to (2). Since Aichholzer et al.~recently proved (1), this
settles (2).  
\end{abstract}

\maketitle

\section{Introduction}

The problem of {\em convexifying} a polygon (that is, continuously
transforming it, while maintaining simplicity) is a classical,
well-studied problem in computational geometry. Different
restrictions on the transformations allowed give rise to several 
versions of this problem. In the most famous variant, 
the lengths of the edges are to be
maintained; this is the celebrated Carpenter's Rule Theorem
(see~\cite{cdr,Stre}).

Devadoss~\cite{demlan} proposed another variant: (1) 
can every polygon be convexified 
without losing internal visibility between any pair of
vertices? (equivalently: does every polygon admit a {\em
  visibility-preserving} convexification?)
 Devadoss~\cite{devadoss} also raised a stronger version of this
 question: (2) does there exist a
 visibility-preserving convexification
in which at most one vertex is moved at a time (that
is, by {\em single-vertex moves})? Partial answers to (1) and (2) were given in~\cite{acflsu} and~\cite{devadoss}.

Our main result is the following.

\begin{theorem}\label{thm:maintheorem}
An affirmative answer to (1) implies an affirmative answer to
(2). That is: if every polygon has a visibility-preserving
convexification, then every polygon has a visibility-preserving convexification using only
single-vertex moves.
\end{theorem}

Since (1) has been recently settled in the afirmative by Aichholzer et
al.~\cite{aetal}, we have the following.

\begin{corollary}
Every polygon has a visibility-preserving convexification using only
single-vertex moves. \hfill$\Box$
\end{corollary}

In Section~\ref{sec:general} we formalize the relevant notions at play
 (such as visibility-preserving transformations and single-vertex
moves). 
The proof of Theorem~\ref{thm:maintheorem} is in
Section~\ref{sec:proofmain}.

\section{Visibility, transformations, and
  single-vertex moves}\label{sec:general}

A {\em polygon} consists of a finite set $\{p_1, p_2,
\ldots, p_n\}$ of points (the {\em vertices}), plus the straight segments (the
      {\em edges}) joining $p_i$ to $p_{i+1}$ for $i=1,2,\ldots,n$
      (indices are read modulo $n$). 
Throughout this work we implicitly assume
      that all polygons under consideration are {\em simple}: edges
      only intersect in a common endpoint.
Two points 
in $P$ are {\em internally visible} (for brevity, {\em visible}) in $P$
if the open segment 
joining them is contained in the
interior of $P$. We also say that the points are $P$-{\em visible}.
If $p,q$ are $P$-visible vertices, then $\{p,q\}$ is a
$P$-{\em visible pair}. 

To speak meaningfully about the continuity of the transformations
under consideration, first we must agree on the underlying metric. Let
$\eucl{p}{q}$ denote the usual Euclidean distance between points $p$ and $q$.  Let
$P,Q$ be polygons, and let $p_1, p_2,
\ldots, p_n$ and $q_1, q_2,\ldots, q_n$ be their respective vertices
(distance is only defined between polygons with the same number of vertices). 
The  {\em
  distance} $d(P,Q)$ between $P$ and $Q$ is $\min_\sigma\{ \max_{i\in\{1,2,\ldots,n\}}\{\eucl{p_i}{
q_{\sigma(i)}} \}  \}$, where the minimum
is taken over all bijections $\sigma:\{1,2,\ldots,n\}\to \{1,2,\ldots,n\}$.

Let
$\pp_n$ denote the collection of all polygons on $n$ vertices. 
If $P\in\pp_n$ and $\delta > 0$, we let
$N_\delta(P)$ denote the set of polygons $Q\in\pp_n$ such that $d(P,Q)
< \delta$.

The polygonal transformations under consideration are obtained by
moving the vertices of the polygon; their incident edges get dragged
along.  Thus, a transformation on a polygon $P$ gets induced by defining, for each
vertex $p$ of $P$, a (time-parameterized) mapping that follows the
movement of
$p$.  The time domain is any closed interval $[a,d]$, and the {\em
  orbit} of $p$ is a continuous mapping $t\mapsto p^t$ such that $p^a
= p$. 
Thus, for each $t\in [a,d]$ we
obtain a polygon $P^t$ (where $P^a = P$), and the transformation
(which we denote $P^{[a,d]}$) is a continuous mapping from $[a,d]$ to
$\pp_n$, transforming the
{\em initial} polygon $P^a =P$ onto the {\em final} polygon $P^d$. 
If $P^d$ is convex, then the transformation is a {\em
  convexification} of $P^a=P$. 
If only one orbit is not the constant map, then the
transformation is a {\em single-vertex move}.

Now let $P\in \pp_n$ be a polygon with vertices $p_1, p_2, \ldots, p_n$. A
transformation $\Pp{a}{d}$ on $P$ is
{\em visibility-preserving} if for all $i,j\in
\{1,2,\ldots,n\}$ and all $a \le s \le t \le d$, if $p^s_i$ and
$p^s_j$ are $P^s$-visible,  then $p^{t}_i$ and $p^{t}_j$ are
$P^{t}$-visible. If in addition there exist $i,j\in \{1,2,$ $\ldots,n\}$ such that $p^a_i$ is not
  $P^a$-visible from $p^a_j$, but $p^d_i$ is $P^d$-visible from
  $p^d_j$, then the transformation is {\em visibility-increasing}.

Let $k\ge 3$. A $k$-tuple $(q_1, q_2, \ldots, q_k)$ of vertices of a polygon $P$ is {\em critical} if 
there is a straight segment that contains $q_1, q_2, \ldots, q_k$, and does not
contain any point in the exterior of $P$. A polygon is {\em
  critical} if it contains a critical $k$-tuple for some $k \ge 3$.
The proof of the following is a trivial exercise. 

\begin{observation}\label{obs:sta}
If $P$ is a critical polygon, then
there is a visibility-increasing single-vertex move with
initial polygon $P$.\hfill$\Box$
\end{observation}

Let $P$ be a polygon, and let $\delta>0$ be small enough so that for
every $Q\in N_\delta(P)$, there is a unique natural bijection between
the vertices of $P$ and the vertices of $Q$. We say that
$\delta$ is a {\em safe radius} for $P$ if for every $Q\in N_\delta(P)$, 
a triple of vertices of $Q$ is critical only if the corresponding triple in $P$ is
also critical.  The proof of the following is straightforward.

\begin{observation}\label{obs:safe}
Every polygon has a safe radius.\hfill$\Box$
\end{observation}

\section{Proof of Theorem~\ref{thm:maintheorem}}\label{sec:proofmain}

Let $P\in\pp_n$, and let $p_1, p_2, \ldots, p_n$ be the vertices of $P$.
With $n$ fixed, we proceed by induction on the number of nonvisible pairs. Since
adjacent vertices are not visible to each other, it follows that the
minimum number of nonvisible pairs is $n$.  If there
are exactly $n$ nonvisible pairs, then $P$ is convex, and there is nothing to
prove. Thus we assume that $P$ has $m>n$ nonvisible
pairs, and that the result holds for every polygon in $\pp_n$
with fewer than $m$ nonvisible pairs.  

By assumption, there exists a visibility-preserving
convexification $\Pp{a}{d}$ of $P$.
An elementary real
analysis argument shows that there exists a 
  smallest $c\in[a,d]$ such that $P^c$ is critical. We note that
  if $c=a$, then by Observation~\ref{obs:sta} 
we are done, using
  the induction hypothesis. Thus we may assume that $c > a$. 

Since $t\mapsto P^t$ is continuous for all $t\in[a,c]$,
Observation~\ref{obs:safe} and the compactness of
$[a,c]$ imply that there is a $\delta>0$ that is a safe radius for $P^t$ for
every $t\in[a,c]$.
A similar elementary continuity argument using the compactness of
$[a,c]$ shows that there exists a $\tau>0$ such that, for all
$s,t\in[a,c]$ such that $|s-t|<\tau$, $d(P^{s},P^{t})<\delta$.

Evidently, we may choose $\tau$
so that there is an integer $L$ such that $c-a=L\tau$.

Let $b:=c-\tau$. We claim that for $\ell=0,1,\ldots, L-2$ there is a sequence of
visibility-preserving single-vertex moves that takes $P^{a+\ell\tau}$ to
$P^{a+(\ell+1)\tau}$. Indeed, it suffices to move (in any
order), for
$i=1,2,\ldots,n$, 
$p_i^{a+\ell\tau}$ to $p_i^{a+(\ell+1)\tau}$ following the orbit $p_i^{[a+\ell\tau,a+(\ell+1)\tau]}$.
The choice of $\tau$ guarantees that each of these moves is 
visibility-preserving.  We conclude that there is a sequence of
visibility-preserving single-vertex moves that takes $P^a$ to $P^b$. 

Once we reach $P^{b}$ we proceed similarly towards $P^c$, moving each
vertex $p^b_i$ towards $p^c_i$, one at a time, for $i=1,2,\ldots,n$.
Consider the first critical polygon $Q$ reached in this last stage
(such a $Q$ obviously exists, since $P^c$ itself is
critical). Visibility losses (respectively, gains) may occur only when
(respectively, immediately after) reaching critical polygons.  Thus,
the choice of $Q$ implies that before reaching $Q$, we cannot have
lost visibility. 

Suppose first that in the moment we reach $Q$ we lose visibility; we
shall derive a contradiction. 
In this case, the moving vertex (we recall we reach $Q$ via a
single-vertex move) must be part of a critical triple at
$Q$, and two of these vertices must lose visibility at the moment of
reaching $Q$. Immediately before reaching $Q$, the corresponding three
vertices clearly satisfy the following (which we call {\em Property A}): each two of them are either a visible
pair or adjacent. Since before reaching $Q$
there is no loss of visibility, the corresponding three vertices in
$P^b$ (and hence all the way back to $P^a$) must also satisfy Property A. 
But the corresponding three vertices in $P^c$ cannot satisfy
Property A: indeed, the choice of $\delta$ and $\tau$, plus the fact that they
are a critical triple in $Q$, imply that they
must form a critical triple in $P^c$, but no critical triple satisfies
Property A. This implies that there is a pair of vertices that are
$P^b$-visible but not $P^c$-visible, contradicting that $\Pp{a}{c}$ is
visibility-preserving.  

Suppose finally that at the moment of reaching $Q$ we do not
lose visibility.  In this case, we have then reached $Q$ from
$P^b$ (and hence from $P^a$ as well) by a sequence of
visibility-preserving single-vertex moves. Since $Q$ is critical, then
we are done by Observation~\ref{obs:sta} and the
induction hypothesis. 
\hfill$\Box$

\end{document}